\documentclass[twocolumn,showpacs,aps,prl,superscriptaddress,floatfix]{revtex4}    

\usepackage{graphicx}
\usepackage{dcolumn}
\usepackage{amsmath}

\input pubboard/babarsym

\newcommand{\BABARPubYear}    {07}
\newcommand{\BABARPubNumber}  {002}
          
\newcommand{\SLACPubNumber}   {12389}

\def\emu {\ensuremath{e^{\pm}\mu^{\mp}}\xspace}
\def\modepill {\ensuremath{B\to \pi\ellell}\xspace}

\def\modepiem {\ensuremath{B\to \pi\emu}\xspace}
\def\modepichll {\ensuremath{B^+\rightarrow \pi^+ \ellell}\xspace}

\def\modepichee {\ensuremath{B^+\rightarrow \pi^+ \epem}\xspace}
\def\modepichmm {\ensuremath{B^+\rightarrow \pi^+ \mumu}\xspace}
\def\modepichem {\ensuremath{B^+\rightarrow \pi^+ \emu}\xspace}
\def\modepizee {\ensuremath{B^0\rightarrow \pi^0 \epem}\xspace}
\def\modepizmm {\ensuremath{B^0\rightarrow \pi^0 \mumu}\xspace}
\def\modepizll {\ensuremath{B^0\rightarrow \pi^0 \ellell}\xspace}
\def\modepizem {\ensuremath{B^0\rightarrow \pi^0 \emu}\xspace}

\def\figurebox#1#2#3{%
    \def\arg{#3}%
    \ifx\arg\empty
    {\hfill\vbox{\hsize#2\hrule\hbox to #2{\vrule\hfill\vbox to #1{\hsize#2\vfill}\vrule}\hrule}\hfill}%
    \else
    {\hfill\epsfbox{#3}\hfill}%
    \fi}
 
\long\def\inst#1{\par\nobreak\kern 4pt\nobreak
    {\it #1}\par\vskip 10pt plus 3pt minus 3pt}

\begin{document}

\begin{flushleft}
  \babar-PUB-\BABARPubYear/\BABARPubNumber\\
  SLAC-PUB-\SLACPubNumber\\
\end{flushleft}

\title{
 Search for the Rare Decay
 \boldmath $B \to \pi \ell^+ \ell^-$
}
%
\author{B.~Aubert}
\author{M.~Bona}
\author{D.~Boutigny}
\author{Y.~Karyotakis}
\author{J.~P.~Lees}
\author{V.~Poireau}
\author{X.~Prudent}
\author{V.~Tisserand}
\author{A.~Zghiche}
\affiliation{Laboratoire de Physique des Particules, IN2P3/CNRS et Universit\'e de Savoie, F-74941 Annecy-Le-Vieux, France }
\author{J.~Garra~Tico}
\author{E.~Grauges}
\affiliation{Universitat de Barcelona, Facultat de Fisica, Departament ECM, E-08028 Barcelona, Spain }
% ...
\collaboration{The \babar\ Collaboration}
\noaffiliation

\date{\today}

\begin{abstract}

We have performed a search for the flavor-changing neutral-current
decays \modepill, where \ellell is either $e^+e^-$ or $\mu^+\mu^-$,
using a sample of $230 \times 10^6$ \FourS $\to$ \BB decays collected with
the \babar\, detector. We observe no evidence of a signal and measure
the upper limit on the isospin--averaged branching fraction to be
${\cal B}(\modepill) < 9.1 \times 10^{-8}$ at 90\% confidence level.
We also search for the lepton-flavor--violating decays \modepiem and
measure an upper limit on the isospin-averaged branching fraction of
${\cal B}(\modepiem) < 9.2 \times 10^{-8}$ at 90\% confidence level.

\end{abstract}
 
\pacs{13.20 He}
\maketitle

\par    
   

In the Standard Model (SM), the decays $B \rightarrow
\pi\ell^+\ell^-$, where \ellell is either $e^+e^-$ or $\mu^+\mu^-$,
proceed through $b \rightarrow d \ell^+\ell^-$ flavor-changing
neutral-current processes (FCNC) that do not occur at tree level.
Three amplitudes contribute at leading order: a photon penguin, a $Z$
penguin, and a $W^+W^-$ box diagram.  With highly suppressed SM rates,
predicted to be $(3.3\pm1.0)\times 10^{-8}$~\cite{bib:aliev1998},
these decays provide a promising means to search for effects of new
flavor-changing interactions.  Such effects are predicted in a wide
variety of models, usually in the context of $b \rightarrow
s\ell^+\ell^-$ decays~\cite{bib:slltheory,bib:TheoryA,bib:lq}.  The
$b\to d \ellell$ decay involves quark-flavor transitions different
from $b \rightarrow s\ell^+\ell^-$ and thus its measurement
constitutes an independent test for new flavor-changing interactions.
An experimentally similar but otherwise unrelated process, the
lepton-flavor--violating (LFV) decay $B \to \pi e^{\pm} \mu^{\mp}$, is
forbidden in the SM, but can occur in some models beyond the SM, such
as theories involving leptoquarks~\cite{bib:lq}. Earlier searches by
other experiments ~\cite{bib:status} have reached branching fraction
upper limits at the $10^{-3}$ level for the FCNC decay and the
$10^{-6}$ level for the LFV decays.


In this Letter we report the findings of a search for the decays
\modepill and \modepiem in 208.9 \invfb\ of data recorded at the
$\Upsilon(4S)$ resonance, corresponding to $(230.1\pm2.5)\times 10^6$
$\BB$ decays.
The data were collected with the \babar\ detector~\cite{ref:babar}
at the \pep2\ storage ring located at the
Stanford Linear Accelerator Center. 
The event selection criteria are optimized using simulated data
and data samples independent of those selected as signal.
The signal model used for efficiency evaluation of the $\ell^+\ell^-$
modes uses form-factors from~\cite{ballzwicky05} and 
amplitudes from~\cite{bib:TheoryA}. 
Calculations of the same type have previously been shown by~\cite{bib:babarkll2006} 
to describe the kinematic distributions of $B\to K\ellell$ well.
The efficiency of the $e\mu$ event selection is estimated using a 3-body
phase space model with QED photon radiative corrections.

We reconstruct signal events by combining two oppositely charged
leptons ($e^+e^-$, $\mu^+\mu^-$ or $e^{\pm}\mu^{\mp}$) with a pion
($\pi^{\pm}$ or $\pi^0$).  Electron (muon) candidates are required to
have a momentum larger than 0.3 (0.7) \gevc. We suppress backgrounds
due to photon conversions in the $B \to \pi\; e^+e^-$ channels by
removing $e^+e^-$ pairs with invariant mass less than 30 \mevcc.
Bremsstrahlung photons from electrons are recovered if the photon has
an energy of $E > 30$ \mev and a direction within a small angular
region around the initial electron momentum vector.  
The identification of electrons (muons) is about 92\% (68\%) efficient 
on average with a hadron misidentification rate of less than 1\% (4\%). 
Charged pion identification is more than 85\% efficient and has a kaon
misidentification rate of less than 5\%. Neutral pions are identified
as pairs of photons, each having an energy of at least 50 \mev. The
invariant mass of the pair is required to satisfy $115 < m_{\gamma
\gamma} < 150 \mevcc$.


Correctly reconstructed $B$ decays produce narrow peaks in the
distributions of two kinematic variables: the beam-energy substituted
mass, $m_{\rm ES} = \sqrt{E_{\rm b}^{*2} - |{\rm p}^*_{B}|^2}$, and
$\Delta E = E_{B}^{*} - E_{\rm b}^*$.  Here, $E_{\rm b}^*$ is the beam
energy and $E_B^*$ ($p_{B}^{*}$) is the energy (momentum) of the
reconstructed $B$ meson, evaluated in
the center-of-mass (c.m.) frame.  For signal events the $m_{\rm ES}$
distribution is centered at the $B$-meson mass and the \DeltaE
distribution is centered at zero. 
The mean and width of these distributions are determined from smearing
and shifting the values from simulated signal events according to
studies of $B^{+} \to J/\psi K^{+}$ and $B^0 \to J/\psi \pi^0$ events
in data control samples and simulations. 
We find the width of \mes to be 2.5 (1.8)
\mevcc for the $\pi^{\pm}$ ($\pi^0$) modes and widths of \DeltaE to be
23, 50, 20 and 39 \mev for the $\pi^{\pm}e^+e^-$, $\pi^0e^+e^-$,
$\pi^{\pm}\mu^+\mu^-$ and $\pi^0\mu^+\mu^-$ final states,
respectively.  For events reconstructed as $e^{\pm}\mu^{\mp}$, we
assume the same mean and width as for the corresponding $e^+e^-$
modes.


The primary sources of background are random combinations of particles
from $e^+e^- \to q\bar q$ ($q = u,d,s,c$) and
from $\Upsilon(4S)\to \BB$ decays. These combinatorial backgrounds
typically arise from pairs of semileptonic decays of $B$ and $D^{(*)}$
mesons. Additionally, there is 
background from events that are peaking in \mes and \DeltaE as they have the 
same topology as signal events. These events include $B\to J/\psi \pi$
($J/\psi\to\ell^+\ell^-$), $B^{\pm}\to J/\psi K^{\pm}$ or $B^{\pm} \to K^{\pm}\ellell$ (with
$K^{\pm}$ misidentified as $\pi^{\pm}$), and $B\to \pi hh$ (with two hadrons $h =
K^{\pm},\pi^{\pm}$ misidentified as muons).


Contributions from $e^+e^- \to q\bar q$ processes are reduced by
exploiting the difference between the spherical track distribution in
$\BB$ events and the jetlike structure of $e^+e^- \to q\bar q$
events. We consider events for which the ratio of second to zeroth Fox-Wolfram
moments $R_2$ is less than 0.5.
Further suppression by a factor of $\sim 45$ is obtained by
constructing a Fisher discriminant 
from the following four quantities~\cite{bib:thesis} defined in the center-of-mass
frame: $R_2$, $|\cos \theta_{thr}|$ where $\theta_{thr}$ is the angle
between the thrust axis of the signal particles and that of the
remaining particles in the event, $|\cos \theta_{B}|$ where $\theta_B$
is the angle of the $B$ candidate's momentum vector with respect to
the beam axis, and the ratio of second- to zeroth-order Legendre
moments~\cite{bib:Legendre}.  

Combinatorial background from $\BB$ events is reduced by a factor of
$\sim 3$ by using a likelihood ratio composed of~\cite{bib:thesis}: the missing energy
of the event (computed from all charged tracks and neutral energy
clusters), the vertex fit probability of all tracks from the $B$
candidate, the vertex fit probability of the two leptons, and $\cos
\theta_{B}$. 
Missing energy provides the strongest suppression of these events,
which typically contain energetic neutrinos from at least two
semileptonic $B$ or $D^{(*)}$ meson decays.


We veto events that have a dilepton invariant mass consistent with the
$J/\psi$ resonance ($2.90 <m_{e^+e^-}< 3.20 \gevcc$ and
$3.00 <m_{\mu^+\mu^-}< 3.20 \gevcc$) or with the $\psi(2S)$ resonance
($3.60 <m_{\ell^+\ell^-}< 3.75 \gevcc$). For electron modes, the vetoes
are applied to $m_{\ellell}$ computed both with and without
bremsstrahlung recovery.  When a lepton radiates or is mismeasured,
$m_{\ell^+\ell^-}$ may shift to values below the charmonium mass, with
$\Delta E$ shifting downward accordingly.
Therefore, we veto events that lie in linearly dependent $\Delta
E$--$m_{\ell\ell}$ bands, whose widths are determined from simulation,
similar to the technique applied in~\cite{bib:babarkll2006}.
For $e^{\pm}\mu^{\mp}$ modes, we use the same vetoes as for the
$e^+e^-$ modes.  In modes with muons, in order to veto events with
tracks that are consistent with hadronic decays $D \to K \pi$ or $D
\to \pi \pi$, we require $m_{\ell \ell}$ and $m_{\pi \ell}$ to lie
outside the range $1.84-1.89$ \gevcc when the $\ell$ is assigned the
mass of a $\pi$ or $K$. For the $\pi^0$ modes, the range for $m_{\pi
\ell}$ is increased to $1.79-1.94$ \gevcc.


The events removed by the charmonium vetoes are kinematically similar
to signal events and serve as large control samples for studying
signal shapes, selection efficiencies, and systematic errors.  The
branching fraction of $B \to J/\psi\, \pi$ is also extracted from
the control sample and found to be in agreement with the current world
average~\cite{PDG}.  We also select a control sample of $B^+ \to
J/\psi K^+$ events to measure the efficiencies and systematic
uncertainties of lepton identification and the Fisher and likelihood
selection.

We extract the signal yield by counting events within a signal region
defined as $\pm 2 \sigma$ around the mean values of the \mes and
\DeltaE distributions expected for signal events, and comparing
observed event counts with estimations of the remaining background in
the same region, summarized in Table~\ref{tab:systematic bkg}.

\begin{table*}[tbp]
\caption{ 
  Number of background events with associated systematic uncertainties 
  expected in the signal region. 
}
\label{tab:systematic bkg}
\begin{center}
\begin{tabular}{lD{,}{\pm}{-1}D{,}{\pm}{-1}D{,}{\pm}{-1}D{,}{\pm}{-1}D{,}{\pm}{-1}D{,}{\pm}{-1}}
 \hline  \hline 
 & \multicolumn{1}{c}{$\pi^+e^+e^-$} & \multicolumn{1}{c}{$\pi^0 e^+e^-$} & \multicolumn{1}{c}{$\pi^+\mu^+\mu^-$} & \multicolumn{1}{c}{$\pi^0 \mu^+\mu^-$} & \multicolumn{1}{c}{$\pi^+ e\mu$} & \multicolumn{1}{c}{$\pi^0 e\mu$} \\ \hline
 \mes -\DeltaE fit  
 & 0.84\ \,,0.24 & 0.43\ \,,0.23 & 0.90\ \,,0.25  & 0.23\ \,,0.20 & 1.55 , 0.34 & 1.22 , 0.43 \\
 \mes -\DeltaE correlations
 & , 0.02    &     ,0.03 &     ,0.06  &     ,0.03 &     ,0.17 &     ,0.05 \\
 \DeltaE shape 
 & , 0.02    &     ,0.01 &     ,0.12  &     ,0.02 &     ,0.31 &     ,0.24 \\ 
 Peaking ($\ell^+\ell^-$) 
 & 0.057,0.016 & 0.009,0.003 & 0.032,0.008 & 0.005,0.001 & 0.0\ \,,0.001\ \,\,\, & 0.0\ \,,0.001\ \,\,\, \\  
 Peaking (hadronic)
 & \multicolumn{1}{c}{$<$ 0.001} & \multicolumn{1}{c}{$<$ 0.001} & 0.027,0.033 & 0.035,0.022 & 0.0\ \,,0.02\ \,\,\, & 0.0\ \,,0.02\ \,\,\, \\ [1ex]
 Total 
 & 0.90\ \,,0.24 & 0.44\ \,,0.23 & 0.96\ \,,0.29  & 0.27\ \,,0.20   & 1.55 , 0.49 & 1.22 , 0.50\\  \hline  \hline
\end{tabular}
\end{center}
\end{table*}

To determine the peaking background from hadronic $B \to \pi\pi\pi$ or
$B \to K\pi\pi$ events, we select a control sample where one track is
required to pass hadron identification in place of muon
identification.  This selects hadronic $B$-decays where 
the remaining track that passes the muon selection is a mis-identified hadron.
Each event is further weighted by the
probability that one more hadron is mis-identified as a muon. The
expected contribution to the $B \to \pi \ellell$ signal region is
extracted from a one-dimensional fit to the distribution of \mes for
events that pass the \DeltaE signal selection.


Backgrounds from $B$ decays to final states with real leptons are
estimated from high-statistics samples of simulated events. $B \to
K\ellell$ events are the largest peaking background component
for the $\pi^+e^+e^-$ mode, but are shifted towards lower \DeltaE than
signal and fall outside the signal region. $B \to \rho\ellell$ events
contribute even less, since the reconstructed $B$ mesons are missing a pion.
Background from charmonium resonances are found to be negligible. 

The expected number of combinatorial background events is extracted
from a two-dimensional, unbinned maximum-likelihood fit to \mes and
\DeltaE in a sideband defined by $5.2 < \mes < 5.2724 \gevcc$
and $|\Delta E| < 0.25 \gev$, i.e., below the \mes value
expected for signal $B$ events.  The signal-region yield is obtained
from extrapolation of this fit into the signal region. 
This procedure has been validated by studies of simulated background
events and data events in the $e\mu$ channel where no signal-like events 
are expected.
The background probability distribution function (PDF) is modeled as
the product of an ARGUS function~\cite{bib:ARGUS} for $m_{\rm ES}$ and
an exponential function for $\Delta E$. The slopes and normalization
are floating in the fit.  Average biases in the background central
value and its uncertainty were corrected for, based on a study of a
large ensemble of simulated experiments generated from the background
PDF obtained from data. The corrections amount to $35\%$ in the
low-statistics \modepizmm channel and $<10\%$ in all others.


Systematic uncertainties due to the background estimates are
summarized in Table~\ref{tab:systematic bkg}.  The uncertainty in the
combinatorial-background estimate is determined by varying the fit
parameters by $\pm 1 \sigma$ of the best fit.
We also consider the effect of using alternative PDF parameterizations
on the background estimates, and use the computed differences to bound
the systematic uncertainty.  Alternatives considered include a PDF
that is correlated in \mes and \DeltaE via a linear \DeltaE dependence
in the \mes slope parameter, and PDFs for which the \DeltaE shape is a
linear or quadratic polynomial.
For peaking background with real leptons the uncertainty is dominated
by limited knowledge~\cite{PDG} of the branching fractions for these
processes, and for hadronic $B$ peaking background the uncertainty is
dominated by the control sample statistics from which it is derived.

Systematic uncertainties due to the signal efficiency include:
charged-particle tracking efficiency (0.8\% per lepton, 1.4\% per
charged hadron) and identification (0.7\% per electron pair, 1.9\% per
muon pair, 0.5\% per pion), neutral pion efficiency (3\%), the Fisher
and likelihood selection (1.4\% for all modes involving electrons,
1.7\% for \modepichmm and 1.9\% for \modepizmm), and signal simulation
statistics (0.1\%).
A systematic uncertainty in signal-region selection efficiency arises
from the uncertainty in the mean and width of the \mes and \DeltaE
distributions determined from charmonium control samples. This
contributes a total uncertainty of 0.7\% for charged modes for which a
high-statistics sample of $B^+ \to J/\psi K^+$ events is used, and a
total of 7\% uncertainty for neutral modes for which a small
statistics sample of $B^0 \to J/\psi \pi^0$ events is used.
For the electron modes, we vary the amount of the bremsstrahlung tail
in the \DeltaE distribution, introducing a systematic uncertainty of
1-1.4\%.
The number of $\BB$ events in the data sample is known to a precision
of 1.1\%.
Additional systematic uncertainties for the efficiency result from the
choice of the form factor model and the relative magnitudes of the
$b\to d\ellell$ amplitudes, which affect the distribution of
four-momentum transfer $q^2 = m_{\ellell}^2$ of the signal.  We
evaluate these systematics from the spread in efficiencies when using
alternative form-factor models~\cite{alternativeff}, and when varying
the Wilson coefficients in the amplitudes by a factor of $\pm2$.  
The former uncertainty varies from 1.1\% for \modepichee to 7.3\% for 
\modepizmm; the latter uncertainty varies from 0.3\% for \modepizmm 
to 1.2\% for \modepichee.
For the $e\mu$ modes we use the spread in efficiency when applying 
two alternative theoretical models for these decays, which amounts 
to 17\% (19\%) for the $\pi^{\pm (0)}$ mode. 
The total systematic uncertainty of the signal efficiencies are
4\% (9\%), 6\% (11\%) and 17\% (21\%)  for 
$\pi^{\pm (0)}ee$, 
$\pi^{\pm (0)}\mumu$ and 
$\pi^{\pm (0)}e\mu$ modes, respectively. 

\begin{figure}[!hbt]
\begin{center}
  \includegraphics[width=0.5\textwidth]{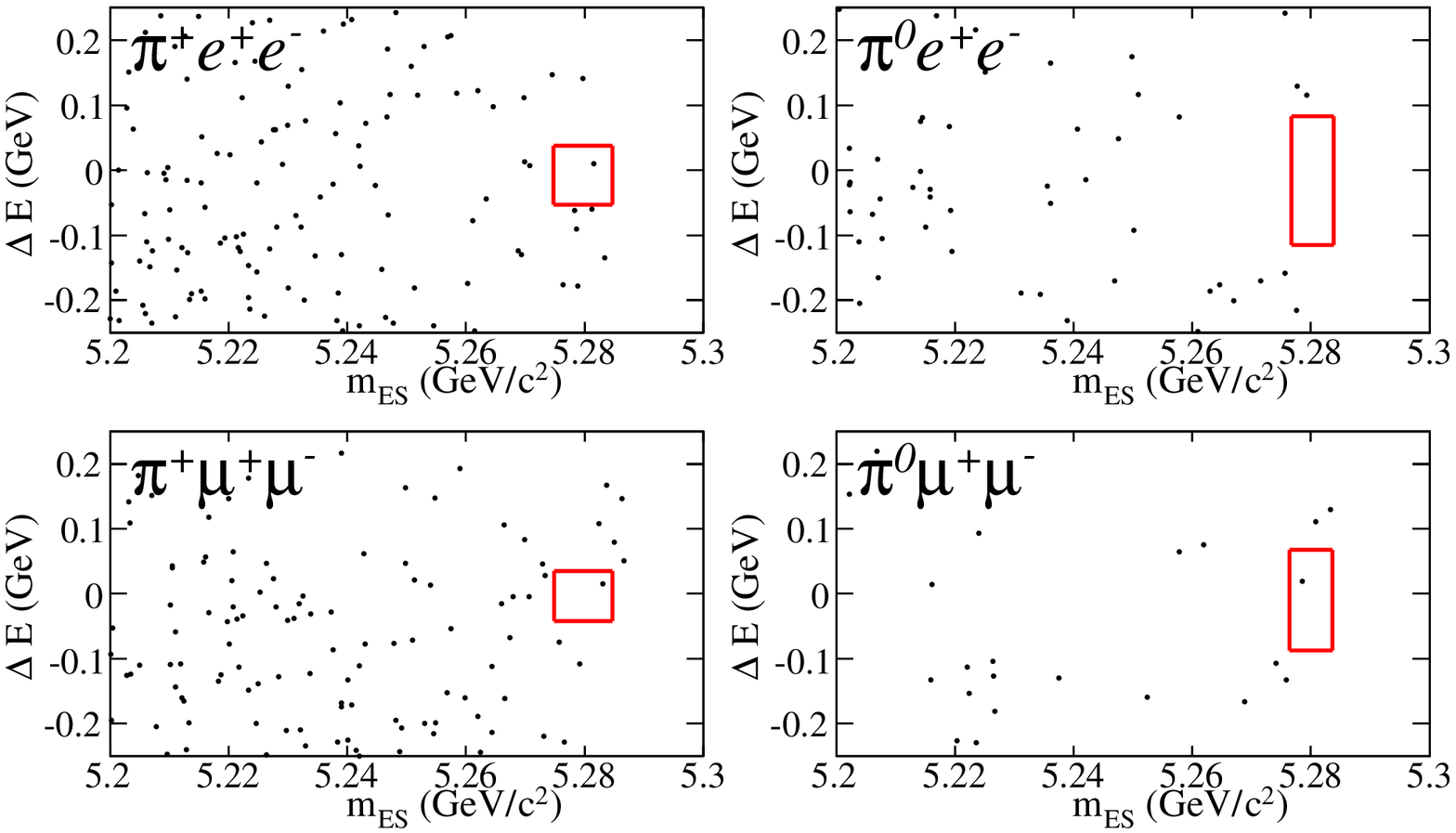}\\
  \includegraphics[width=0.5\textwidth]{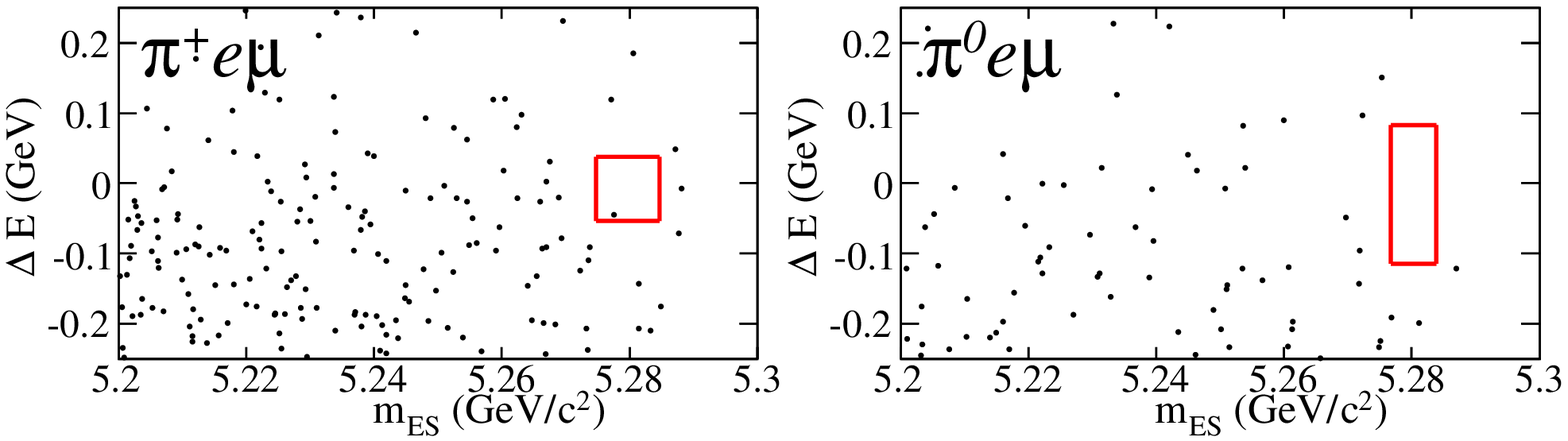}
  \caption{
    $m_{\rm ES}$--$\Delta E$ distributions for events selected in each mode.  
    The rectangles indicate the signal regions.
    \label{fig:scatter}
  }
\end{center}
\end{figure}

Figure~\ref{fig:scatter} shows the distribution of events from data in
the \mes--\DeltaE plane. The rectangles in the plots indicate the
signal regions. Three \modepill candidates and one \modepiem candidate
are observed in the signal regions, which is consistent with the
expected background.  In Table~\ref{tab:results} we calculate the
branching fraction upper limits at 90\% confidence level (C.L.) using a
frequentist method that takes systematic uncertainties and their
correlations into account. We follow the algorithm
of~\cite{bib:barlow}, but differ from it in that we assume Gaussian 
distribution truncated at zero for the systematic uncertainties in signal
sensitivity and background expectation. 
We combine modes and determine the
$e$-$\mu$--averaged branching fractions to be ${\cal B}(\modepichll) <
1.2 \times 10^{-7}$ and ${\cal B}(\modepizll) < 1.2 \times 10^{-7}$ at
90\% C.L., where charged conjugate modes are implied. 
Defining the isospin averaged branching fraction  ${\cal B}(\modepill) \equiv 
{\cal B}(\modepichll) = 2 \times \frac{\tau_{B^+} }{ \tau_{B^0} } {\cal B}(\modepizll)$,
where the different $B$-meson lifetimes $\tau_B$~\cite{PDG} are taken into account, 
we find the combined upper limit
\begin{center}
${\cal B}(\modepill) < 9.1\times10^{-8}$ at 90\% C.L.
\end{center}
This is about a factor three above the nominal SM prediction~\cite{bib:aliev1998}.
We similarly compute the combined limit for the $e\mu$ modes of 
\begin{center}
${\cal B}(\modepiem) < 9.2\times 10^{-8}$ at 90\% C.L.
\end{center}

\begin{table}[tbp!]
 \caption{
   The observed yields, number of expected
   background events, signal efficiency, and branching fraction
   (${\cal B}$) upper limit (U.L.) at 90\% C.L. in
   units of $10^{-7}$. The upper limits for combined modes are also given.
 }
 \label{tab:results}
 \begin{center}
 \begin{tabular}{lccccc}
   \hline\hline
   &\multicolumn{1}{c}{Observed} 
   &\multicolumn{1}{c}{Expected}
   &\multicolumn{1}{c}{Signal}
   &\multicolumn{1}{c}{${\cal B}$ U.L.}\\
   \multicolumn{1}{c}{Mode}   
   &\multicolumn{1}{c}{events} 
   &\multicolumn{1}{c}{background}
   &\multicolumn{1}{c}{efficiency} 
   &\multicolumn{1}{c}{90\% C.L.} \\
   \hline 
   \modepichee         &$1$&  $0.90\pm0.24$& $(7.1\pm0.3)\%$ & $1.8~$ \\
   \modepizee          &$0$&  $0.44\pm0.23$& $(5.7\pm0.5)\%$ & $1.4~$ \\
   \modepichmm         &$1$&  $0.96\pm0.29$& $(4.7\pm0.3)\%$ & $2.8~$ \\
   \modepizmm          &$1$&  $0.27\pm0.20$& $(3.1\pm0.3)\%$ & $5.1~$ \\
   \modepichem         &$1$&  $1.55\pm0.49$& $(6.3\pm1.1)\%$ & $1.7~$ \\
   \modepizem          &$0$&  $1.22\pm0.50$& $(3.7\pm0.8)\%$ & $1.4~$ \\ [1ex]

   \modepichll         &&&                                & $1.2~$ \\
   \modepizll          &&&                                & $1.2~$ \\
   \modepill           &&&                                & $0.91$ \\
   \modepiem           &&&                                & $0.92$ \\
   \hline\hline
 \end{tabular}
 \end{center}
\end{table}

In conclusion, we have presented the result of a search for $B \to \pi
\ell^+\ell^-$ using a sample of $(230.1\pm2.5)\times 10^6$ $\BB$ pairs
produced at the \FourS resonance. 
No excess of events is observed in the signal regions, and at $90\%$
confidence limit we measure the upper limit of ${\cal B}(\modepill) <
9.1 \times 10^{-8}$, which is within a factor three of SM
expectations.  We also measure the upper limit of the lepton-flavor--violating
branching fractions to be ${\cal B}(\modepiem) < 9.2 \times 10^{-8}$.


We are grateful for the excellent luminosity and machine conditions
provided by our \pep2\ colleagues, 
and for the substantial dedicated effort from
the computing organizations that support \babar.
The collaborating institutions wish to thank 
SLAC for its support and kind hospitality. 
This work is supported by
DOE
and NSF (USA),
NSERC (Canada),
CEA and
CNRS-IN2P3
(France),
BMBF and DFG
(Germany),
INFN (Italy),
FOM (The Netherlands),
NFR (Norway),
MIST (Russia),
MEC (Spain), and
STFC (United Kingdom). 
Individuals have received support from the
Marie Curie EIF (European Union) and
the A.~P.~Sloan Foundation.



\begin{thebibliography}{99}
\bibitem{bib:Legendre}{The $i^{\mathrm{th}}$ moment $L_i$ is defined by
$L_i = \sum_j |p^*_j| |\cos(\theta^*_j)|^i$,
where the $p^*_j$ are the c.m.\ momenta of all particles not used in
reconstructing the signal $B$ candidate, and the angle $\theta^*_j$ is
between the particle's momentum and the thrust axis of the signal
$B$. }
\end{thebibliography}
\end{document}